\documentclass[osajnl2,twocolumn,showpacs]{revtex4}

\usepackage{hyperref}
\usepackage{amsmath}
\usepackage{calc}
\usepackage{amsfonts}
\usepackage{amssymb}
\usepackage{graphicx}
\usepackage{bm}
\usepackage{color}
\usepackage{epsfig}

\usepackage{verbatim}
\usepackage{ulem}

\def\jones{\mathbf{J}}
\def\ret0{\varphi_0}

\begin{document}

\title{Variable Ultra-broadband and Narrowband Composite Polarization Retarders}

\author{Thorsten Peters$^{1,*}$, Svetoslav S. Ivanov$^{2,\dagger}$, Daniel Englisch$^{1}$, Andon A. Rangelov$^{2}$, Nikolay V. Vitanov$^{2}$ and Thomas Halfmann$^{1}$}

\address{
$^1$Institut f\"{u}r Angewandte Physik, Technische Universit\"{a}t Darmstadt, Hochschulstrasse 6, 64289 Darmstadt, Germany\\
$^2$Department of Physics, Sofia University, James Bourchier 5 blvd., 1164 Sofia, Bulgaria\\
$^*$Corresponding author: thorsten.peters@physik.tu-darmstadt.de\\
$^\dagger$Corresponding author: sivanov@phys.uni-sofia.bg}

\begin{abstract}
We propose and experimentally demonstrate novel types of composite sequences of half-wave and quarter-wave polarization retarders, either permitting operation at ultra-broad spectral bandwidth or narrow bandwidth. The retarders are composed of stacked standard half-wave retarders and quarter-wave retarders of equal thickness. To our knowledge, these home-built devices outperform all commercially available compound retarders, made of several birefringent materials.
\end{abstract}


\maketitle

\section{Introduction}

Optical half-wave and quarter-wave retarders \cite{Hecht} based on birefringent materials are standard optical components with innumerous applications.
Light passing through a birefringent material with the optical axis aligned perpendicular to the beam propagation direction
experiences a phase shift $\varphi=2\pi/\lambda(n_e-n_o)L$ between the two orthogonal polarization directions.
Here, $\lambda$ is the light wavelength, $L$ is the thickness of the birefringent material and $n_{e,o}$
are the refractive indices of the (extra-)ordinary rays. The most common types of retarders are half-wave plates (HWPs)
with $\varphi=(n+1)\pi$ and quarter-wave plates (QWPs) with $\varphi=(n+1/2)\pi$, where $n=0,1,2,...$.
These wave plates are divided into multi-order wave plates (with $n>0$) and zero-order wave plates (with $n=0$).
In the optical region, zero-order wave plates are typically constructed by combining two multi-order plates with
crossed optical axes to yield the desired phase shift. A single zero-order wave plate would be very thin and too fragile.

Both zero- and multi-order wave plates show dispersion, i.e., the phase shift (retardation) depends on the wavelength of the incident light. This effect is more pronounced for multi-order than for zero-order plates. However, many applications require achromatic (broadband, BB) wave plates, which show negligible dispersion over a large wavelength range. As examples for applications of BB plates we note, e.g., terahertz (THz) time-domain spectroscopy
\cite{GKE90,MG06}, microwave polarimetry \cite{HHJ05,PSA06,MHA09}, or achromatic polarization rotators in liquid crystal devices \cite{ZKP00,LSK04}.
There are several types of achromatic wave plates. \textit{Compound}-type achromatic wave plates utilize two different materials. If the wavelength-dependent
difference of the refractive indices $\Delta n(\lambda) = n_e(\lambda)-n_o(\lambda)$ of the two materials (almost) cancel each other, the compound wave plate is
achromatic over a larger wavelength range \cite{D67,B71,B72}. \textit{Stacked}-type achromatic wave plates use a stack of (conventional) wave plates
\cite{P55b,K59,T75,MH68,LSK04}. Here, the optical axis of each wave plate is rotated with respect to the other plates to create a specific sequence of phase shifts.
Appropriate choice of the phase shift sequences permits (almost) achromatic retardation over a large bandwidth or for a larger set of discrete wavelengths.

Recently, Ardavan pointed out \cite{A07b} that states from the Poincar$\acute{\mathrm{e}}$ space
(which are typically used for the description of photon polarizations) can be mapped onto the Bloch space (which is typically used for the description
of quantum state dynamics of two level systems, e.g. spin $1/2$ particles). This allows transfer of concepts from the field of coherent interactions between light and quantum systems to the field of polarization optics. In particular, we draw our attention now to techniques for error correction during quantum state transfer \cite{W94,CLJ03}. The latter are widely used in the field of nuclear magnetic resonance (NMR). The concepts use a sequence of composite pulses \cite{L86} to drive a quantum system, e.g., from a ground to an excited state.
By appropriate choice of the phases in the composite pulses, the sequence is fault-tolerant. Thus, the efficiency of the transfer from a ground to an excited quantum state does not change with variations in the pulse parameters (e.g., intensity or frequency). In terms of the dynamics (i.e., rotation) of the Bloch vector in Bloch space, composite pulses are fault-tolerant with respect to rotation errors induced by individual pulses.

Based on this concept of composite pulses for rotations in the Bloch space, Ardavan proposed to use the so-called \textit{BB1} or \textit{BB2} sequences \cite{W94} for polarization retarders
(i.e., rotations in Poincar$\acute{\mathrm{e}}$ space). Ardavan already found, that these \textit{stacked} composite retarders in almost all cases
outperform the conventional \textit{compound}-type retarders \cite{A07b}. Recently, Ivanov \textit{et al.} applied a systematic theoretical approach to obtain even better sequences of composite retarders which outperform the \textit{BB1}-related sequences in terms of bandwidth \cite{IRV12}.

In contrast to achromatic retarders, other applications require highly chromatic (narrowband, NB) polarization retarders. As applications we note Lyot \cite{L44} or $\check{\mathrm{S}}$olc filters \cite{S53}, e.g., for solar imaging at specific narrow wavelength bands \cite{BDJ75,KDE97}.
The $\check{\mathrm{S}}$olc filter is quite reminiscent of a narrowband stacked retarder. It consists of several wave plates placed between two polarizers. Both the Lyot and the $\check{\mathrm{S}}$olc filter suffer from residual transmission sidebands outside the center transmission window \cite{E58,S65}. This requires improvement. Also here, transfer of concepts from the background of composite pulses will permit the implementation of novel sequences of wave plates (as we demonstrate below).

In our following work, we present new composite sequences for broadband and narrowband half-wave retarders as well as quarter-wave retarders. The new sequences outperform almost all previous
(i) compound and composite retarders for broadband operation and (ii) composite retarders for narrowband operation. Beyond our recent work on related subjects \cite{IRV12},
we will also present sequences which show non-perfect retardation to a certain degree, in order to increase (or decrease) the bandwidth. In addition, we experimentally verify the performance of the previously proposed \textit{BB1} sequence and several new composite retarders.

We note that for the specific case of THz radiation a stacked achromatic QWP exists \cite{MG06} which works with similar performance as our sequences presented below. This QWP for the THz domain, however, does not consist of standard HWPs and QWPs, but is composed of plates with varying thickness. This is not a very practical approach (at least in the optical frequency domain) due to the need of specially designed (hence, expensive) retarders. On the other hand, in other domains this approach seems to be the most effective, as shown, e.g., by Shaka \cite{S85}, who optimized the detuning bandwidth of the spin inversion for NMR experiments similarly. We further note that according to Clarke \cite{C04c}, the use of \textit{compound} achromatic waveplates for the construction of the \textit{stacked} type, significantly increases the performance. This, however, also results in a much higher cost of the achromatic waveplate sequences.

\section{Theory}\label{sec:theory}

We briefly summarize now the basic theory of composite sequences for optical retarders.
In the Jones calculus \cite{J41}, a single birefringent retarder is described by the matrix
\begin{equation}
\textbf{J}_{\theta}(\varphi) =
R(-\theta)
\left[
\begin{array}{cc}
e^{i\varphi/2} & 0 \\
0 & e^{-i\varphi/2}%
\end{array}%
\right]
R(\theta),
\end{equation}
where
\begin{equation}
R(\theta)=\left[
\begin{array}{cc}
\cos \theta & \sin \theta \\
-\sin \theta & \cos \theta%
\end{array}%
\right].
\end{equation}
We use the linear polarization basis, a pair of orthogonal polarization vectors, where $\theta$ is the rotation angle of the retarder's optical axis and $\varphi$ is the phase shift applied between the ordinary and the extraordinary ray passing through the retarder ($\varphi=\pi$ for HWPs and $\varphi=\pi/2$ for QWPs).
We describe composite retarders (i.e., a stack of $N$ retarders) by the matrix
\begin{equation}
\textbf{J}^{\left( N\right) }=\textbf{J}_{\theta _{N}}\left( \varphi _{N}\right) \textbf{J}_{\theta_{N-1}}\left( \varphi _{N-1}\right) \cdots \textbf{J}_{\theta _{1}}\left( \varphi_{1}\right),
\label{overall Jones matrix}
\end{equation}
where the light passes through the plates described by $\textbf{J}_{\theta _{i}}\left( \varphi _{i}\right)$ in the order of ascending index $i$.
For practical purposes we make two assumptions. First, we take all plates to be either HWP [H] or QWP [Q] with respect to a chosen wavelength $\lambda$ of interest. This allows us to use commercially available, standard wave plates. Second, we choose symmetric configurations of plates with respect to a central plate, where the rotation angles fulfill $\theta_i=\theta_{N-i+1}$. This makes it possible to realize all sequences by a double-pass configuration, where the first half of the sequences are passed twice after reflection at a mirror [M], and by replacing the central HWP by a QWP (see Fig.~\ref{fig:setup} and \cite{IRV12}).
Thus we assume the following configuration of $n$ plates and one mirror
\begin{equation}
[\text{X}_{\theta_{1}}\text{H}_{\theta_{2}} \cdots \text{H}_{\theta_{n-1}} \text{Q}_{\theta_{n}}\text{M}],
\end{equation}
with $\text{X}=[\text{H}]$ for a half-wave retarder and $\mathrm{X}=[\text{Q}]$ for a quarter-wave retarder. This is equivalent to a symmetric sequence of $ N=2n-1$ individual plates.

We denote the Jones matrix of our desired composite retarder by $\jones$.
The (composite) retardation $\delta$ is the difference between the  arguments of the eigenvalues of $\jones$.
Taking into account the angle-symmetry condition of $\mathbf{J}^{(N)}$, we obtain
\begin{equation}
\delta=2\cos^{-1}\text{Re}\jones_{11},
\end{equation}
with $\text{Re}\jones_{11}$ being the real part of the element $\jones_{11}$.
We note that the retardation profile is symmetric relative to the target retardation $\ret0$, i.e., $\delta(\varphi)=\delta(2\ret0-\varphi)$, where $\ret0$ is $\pi$ for a half-wave retarder and $\pi/2$ for a quarter-wave retarder, and $\varphi$ is the phase shift, introduced by the first plate of the sequence. This is because replacing $\varphi$ with $2\ret0-\varphi$ in the Jones matrix $\textbf{J}^{(N)}$ is equivalent to replacing $\theta_i$ with $\theta_i+\pi/2$ for HWPs or with $-\theta_i$ for QWPs. The latter correspond to particular operations of rotation or inversion in space, which produce an identical retarder.

We vary the $n$ rotation angles $\theta_i$ to obtain different broadband and narrowband composite retarders. For \textit{broadband retarders} the angles $\theta_i$ satisfy the condition
\begin{equation}
\text{max}~ \left|\delta(\varphi)-\ret0\right|\leq \Delta, \quad \varphi\in \left[\varphi_\text{min},\ret0\right].
\label{eqn:numeric1}
\end{equation}
This, combined with the symmetry relation $\delta(\varphi)=\delta(2\ret0-\varphi)$, guarantees that there is a range of single plate retardations between $\varphi_\text{min}$ and $2\ret0-\varphi_\text{min}$ (with $\varphi=\ret0$ corresponding to $\lambda$), where the deviation of $\delta$ from $\ret0$ stays below $\Delta$.
For QWPs we explicitly impose $\delta(\ret0)=\ret0$, leaving $n-1$ angles $\theta_i$ free to vary, while for HWPs this is automatically fulfilled. We determine numerically those angles $\theta_i$, which minimize $\varphi_\text{min}$, giving the broadest possible range around $\ret0$ of high-quality retardation. We choose different values for $\Delta$, which define three classes: $\Delta=0.001\pi$ for class I, $\Delta=0.005\pi$ for class II, and $\Delta=0.01\pi$ for class III. This allows us to trade retardation quality for bandwidth. Retarders of class III therefore offer the broadest retardation range.

On the other hand, a \textit{narrowband retarder} must retard only a small region around the wavelength of interest $\lambda$. This imposes the following relations on the angles $\theta_i$:
\begin{equation}
\text{max}~ \left|\delta(\varphi)\right|\leq \Delta, \quad \varphi\in \left[0,\varphi_\text{max}\right],
\label{eqn:numeric2}
\end{equation}
This guarantees that the retardation window, enclosed by $\varphi_\text{max}$ and $2\ret0-\varphi_\text{max}$, is as narrow as possible.
Outside this window, retardation is maintained below $\Delta$.
For QWPs we explicitly impose $\delta(\ret0)=\ret0$, leaving $n-1$ angles $\theta_i$ free to vary, while for HWPs this is automatically fulfilled.
We seek those angles $\theta_i$, which maximize $\varphi_\text{max}$, giving the narrowest possible retardation window around $\ret0$. The three classes defined for broadband retarders apply also to narrowband retarders.

We set the optical axes of the composite retarders at angle 0$^\circ$, which imposes additional constraint on the angles. For HWPs we have $2\sum_{i=1}^{n-1}(-1)^i\theta_i + (-1)^n\theta_n=0,\pm\pi/2$ for BB and $\pm\pi/2$ for NB. For QWPs we have $\theta_1=\pi/4$. 
We use Newton's gradient-based method for the numerical optimization. To determine the optimal solution to Eqns. (\ref{eqn:numeric1}) or (\ref{eqn:numeric2}) out of the many possible, we gradually increase $\varphi_{\text{min}}$ or $\varphi_{\text{max}}$, respectively, which are treated as parameters, until eventually we end up with the optimal set of angles. Because we use a local optimization algorithm, we iteratively pick the initial parameter values using a Monte-Carlo scheme.

The calculated results for the rotation angles $\theta_i$ of the optical axes of the individual plates for several composite sequences are shown in Table~\ref{tbl:data}.
The table shows sequences for broadband and narrowband HWPs as well as QWPs. All sequences consist of an odd number $N$ of plates and are symmetric
with respect to the central plate. Thereby, we can experimentally realize a sequence with effectively $N$ plates using only $n=(N+1)/2$ plates in reality. 
We choose sequences of $N\geq5$, which provides a significant improvement with regard to previously proposed sequences \cite{IRV12}.
\begin{table*}
  \centering
  \caption{Calculated angles (in degrees) of the optical axes of $5\leq N \leq 9$ individual wave plates to implement composite sequences of broadband and narrowband HWPs as well as QWPs.
Due to the symmetry of the sequences, we only give the angles of the first half of plates. The angles of the second half should be reversed. Classes (I,II,III) have a retardation modulation of $(0.001, 0.005, 0.01)\pi$ within (outside) their design bandwidth for the broadband (narrowband) sequences. There is only one narrowband quarter-wave sequence (class I) for $N=5$ due to the low number of free parameters.\label{tbl:data}}
  \begin{tabular}{cccc} \hline\hline
    \multicolumn{4}{c}{(i) broadband half-wave retarders}\\ \hline\hline
    Class & \multicolumn{3}{c}{($\theta_1,..,\theta_{n}$) for plate configurations [HH...HQM]} \\ \hline
     & N=5 (n=3)& N=7 (n=4)& N=9 (n=5)\\
    I & (37.1, 4.1, 114.1) & (51.5, 72.8, 118.2, 12.4) & (138.5, 152.9, 3.5, 54.2, 130.2) \\
    II & (37.7, 7.4, 119.4) & (50.0, 67.9, 108.0, 179.3) & (132.0, 120.5, 95.3, 51.1, 158.5) \\
    III & (38.0, 9.4, 122.8) & (49.2, 65.6, 103.1, 172.7) & (132.2, 121.9, 99.2, 58.5, 168.1) \\ \hline\\
    \hline\hline
    \multicolumn{4}{c}{(ii) broadband quarter-wave retarders}\\ \hline\hline
    Class & \multicolumn{3}{c}{($\theta_1,..,\theta_{n}$) for plate configurations [QH...HQM] } \\ \hline
     & N=5 (n=3)& N=7 (n=4)& N=9 (n=5)\\
    I & (45.0, 94.3, 166.1) & (45.0, 15.7, 110.9, 167.9) & (45.0, 7.7, 151.9, 99.5, 23.1) \\
    II & (45.0, 93.3, 164.1) & (45.0, 84.7, 129.3, 21.5) & (45.0, 81.0, 113.0, 160.3, 53.9) \\
    III & (45.0, 92.6, 162.7) & (45.0, 83.9, 126.4, 17.1) & (45.0, 80.2, 110.0, 154.1, 45.8) \\ \hline \\ \hline\hline
    \multicolumn{4}{c}{(iii) narrowband half-wave retarders}\\ \hline\hline
    Class & \multicolumn{3}{c}{($\theta_1,..,\theta_{n}$) for plate configurations [HH...HQM]} \\ \hline
     & N=5 (n=3)& N=7 (n=4)& N=9 (n=5)\\
    I & (23.3, 98.5, 150.5) & (22.8, 97.7, 154.3, 68.9) & (26.6, 62.7, 159.5, 104.2, 141.5) \\
    II & (23.3, 98.0, 149.4) & (23.0, 97.5, 146.1, 53.1) & (154.3, 123.2, 25.9, 77.1, 40.3) \\
    III & (23.1, 97.3, 148.3) & (23.2, 97.4, 149.6, 60.9) & (139.5, 33.8, 55.3, 113.1, 174.2) \\
\hline\\\hline\hline
    \multicolumn{4}{c}{(iv) narrowband quarter-wave retarders}\\ \hline\hline
    Class & \multicolumn{3}{c}{($\theta_1,..,\theta_{n}$)) for plate configurations [QH...HQM]} \\ \hline
    & N=5 (n=3)& N=7 (n=4)& N=9 (n=5)\\
    I & (45.0, 135, 22.5) & (45.0, 161.3, 101.5, 37.8) & (45.0, 108.7, 168.5,51.1, 140.0) \\
    II & -- & (45.0, 161.9, 102.9, 39.5) & (45.0, 143.5, 89.4, 31.9, 149.7) \\
    III & -- & (45.0, 162.2, 104.0, 41.1) & (45.0, 140.5, 83.5, 26.0, 143.6) \\ \hline
\hline
  \end{tabular}
\end{table*}

\begin{figure}[hbt]
\centerline{
\includegraphics[width=7.0cm]{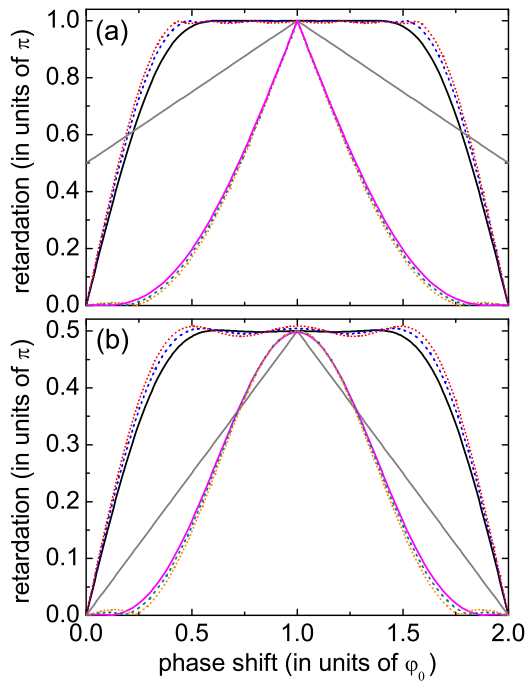}}
 \caption{(Color online) Numerically calculated retardations vs. phase shift for several HWP (a) and QWP sequences (b) with $N=7$. The sequences use the values from Table~\ref{tbl:data}.
Color coding is as follows: conventional single zero-order waveplate (gray); broadband HWPs/QWPs class (I,II,III) (black solid, blue dashed, red dotted); narrowband HWPs/QWPs class (I,II,III)
(magenta solid, dark cyan dashed, orange dotted).}
 \label{fig:theory}
\end{figure}

Figure~\ref{fig:theory} shows calculated optical retardations for several broadband and narrowband sequences with N=7 effective plates vs. phase shift $\varphi/\varphi_0$. Figure~\ref{fig:theory}(a) depicts HWP sequences, whereas Fig.~\ref{fig:theory}(b) depicts QWP sequences. The phase shift $\varphi_0$ is defined with respect to the desired shift at the design wavelength of the individual plates, i.e., $\pi$ for HWPs and $\pi/2$ for QWPs.
The plots for different numbers $N$ of plates look quite similar. A discussion of the effect of $N$ onto the retardation bandwidth can be found in \cite{IRV12}.
For comparison, in each plot we show the retardation of a conventional single plate (zero-order) retarder (gray lines).

\begin{figure}[htb]
\centerline{
\includegraphics[width=7.5cm]{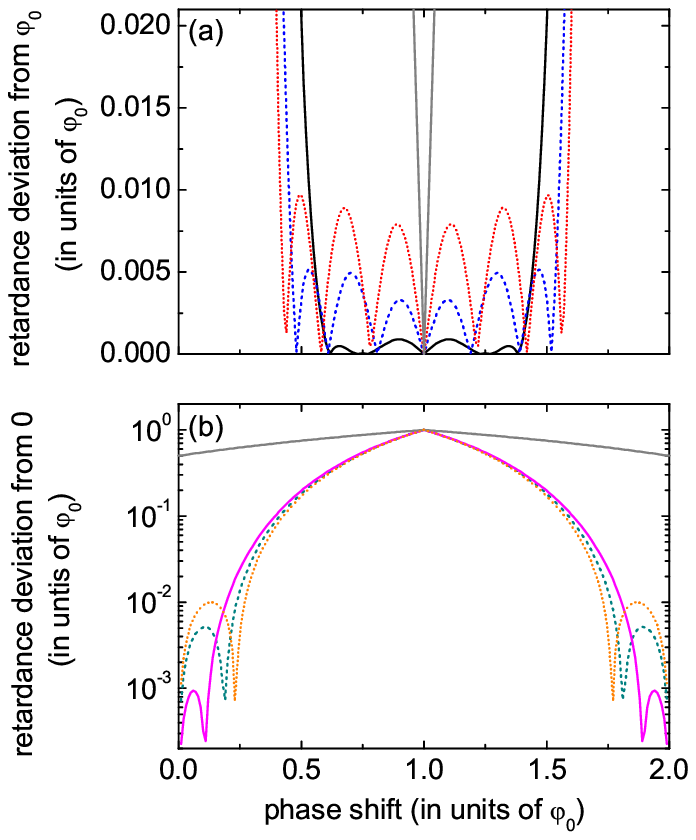}}
 \caption{(Color online) Numerically calculated deviations of the optical retardation for the HWP sequences shown in Fig.~\ref{fig:theory}(a) vs. phase shift.
In (a) we plot the deviation from the target retardation $\varphi_0$ for the broadband sequences.
In (b) we plot the deviation from a retardation of zero for the narrowband sequences. Color coding is as in Fig.~\ref{fig:theory}(a).
For comparison we show the data for a single conventional wave plate (gray lines).}
 \label{fig:error}
\end{figure}

We first discuss the behavior of the broadband sequences (indicated by black solid, blue dashed, and red dotted lines in each plot). Compared to the single conventional plate, the broadband sequences show much less variation of the retardation around the design phase shift $\varphi_0$.
This becomes very clear in Fig.~\ref{fig:error}(a) where we depict the deviation from the target retardation $\varphi_0=\pi$ for the broadband HWPs.
Obviously, the bandwidth of the retarders can be increased by allowing for larger deviations from the target retardation. For example, by allowing a relative deviation of $0.01\pi$ (class III) from a target retardation of $\pi$ instead of $0.001\pi$ (class I), the bandwidth increases from 0.81$\varphi_0$ (see black solid line in Fig.~\ref{fig:theory}(a)) to 1.17$\varphi_0$ (see red dotted line in Fig.~\ref{fig:theory}(a)).
Such an increase of operation bandwidth at the expense of less accuracy of retardation was shown and discussed for three and five stack sequences in \cite{HHJ05}. The feature is in particular important for experimental implementations, which anyway suffer from inevitable losses (e.g., due to non-perfect polarization of light beams or wavelength-dependent reflections). Thus, we can reduce the required number of plates in the composite broadband retarder, while still maintaining a large bandwidth.

We draw now our attention to the behavior of the narrowband sequences, depicted in Fig.~\ref{fig:theory} by the magenta solid, dark cyan dashed, and orange dotted lines.
In contrast to the broadband sequences, the optical retardation changes here much faster than for a single conventional plate. This allows the setup of filters, which change the polarization only in a narrow spectral range (e.g., such as the Lyot \cite{L44} or $\check{\mathrm{S}}$olc filters \cite{S53}). It also permits matching of polarizations in two collinear beams at different wavelengths, even if they were overlapped by a polarizing beamsplitter before. As for the broadband sequences, the bandwidth can also be adjusted for narrowband operation by allowing certain deviations (now from a retardation of zero) for phase shifts away from $\varphi_0$ (see Fig.~\ref{fig:error}(b)). Similar dependencies also show up for QWPs.

We note, that the phase shift $\varphi$ relates to the wavelength $\lambda$ via $\Delta\varphi=2\pi\Delta n(\lambda)L/\lambda$,
where $\Delta n(\lambda) = n_o(\lambda) - n_e(\lambda)$ is the wavelength-dependent difference in the refractive index for the ordinary and extraordinary waves.
For operation at ultra-broad spectral bandwidth, we must consider the dependence $n_{o,e}(\lambda)$ for realistic predictions (see Sec.~\ref{sec:results}).

To the best of our knowledge, our sequences for composite broadband and narrowband retarders made of identical individual plates are the most efficient sequences (in terms of bandwidth) discussed in the literature so far. In all previous works, we find only one composite achromatic QWP for the THz range which shows a similar performance (for a higher number of free parameters)\cite{MG06}. However, this particular composite QWP is made of non-standard retarders (i.e., with $\varphi_0\neq\pi,\pi/2$) and therefore is rather difficult to implement experimentally in the optical region.

\section{Experiment}\label{sec:experiment}

We performed experiments to verify our predictions for the composite optical retarders. In these measurements we used up to 5 individual commercially available zero-order waveplates (\textit{Casix}, WPZ1225-L/2-780 and WPZ1225-L/4-780, 1~inch diameter). These waveplates were designed and anti-reflection (AR) coated for a center wavelength of 780~nm. We determined the optical axes of the plates with an accuracy of $\pm1^\circ$.
Manual rotation mounts held each plate, which allowed us to adjust the relative angles of the optical axis and tilt the plates with respect to each other. Tilting the plates by a few degrees turned out to be necessary due to the small AR coating bandwidth as compared to the ultra-broad retardation bandwidths of the composite sequences. Otherwise, outside the bandwidth of the AR coating, reflections at the interfaces would lead to numerous reflexes which would prevent a measurement of the conversion efficiency without separating the reflexes from the main beam. Also, the resulting interference would be detrimental to the performance of the retarders \cite{C04c}. We note that tilting the plates leads to a shift of the central wavelength. For small angles (as in our experiments) this shift is negligible.

We applied several distinct light sources (i.e., lasers) to demonstrate the ultra-broad and narrow bandwidths of the composite sequences. Standard laser diodes provided radiation at 405~nm, 532~nm, 780~nm, and 850~nm. A Helium-Neon laser served as a light source at 633~nm.
A fiber laser-pumped (\textit{IPG Photonics} Model YAR-15K-1064-LP-SF) optical parametric oscillator (\textit{Lockheed Martin Aculight} Argos Model 2400)
provided radiation at 1550~nm (signal) and at 1064~nm (pump).

For an exemplary experimental demonstration of the composite sequences we restricted ourselves mainly to symmetric sequences with an effective number
of $N=7$ plates for the HWPs and $N=5$ plates for the QWPs. Apart from this, we also tested the asymmetric \textit{BB1} sequence \cite{W94}, previously proposed
by Ardavan \cite{A07b}. The tests worked as follows (see also Fig.~\ref{fig:setup}):
\begin{figure}[htb]
\centerline{
\includegraphics[width=8.5cm]{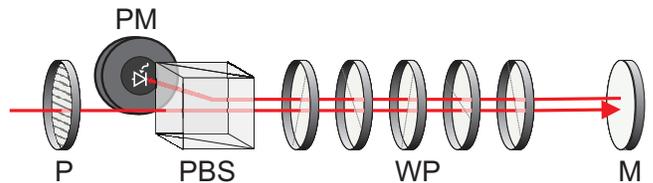}}
 \caption{(Color online) Experimental setup. P: polarizer, PM: powermeter, PBS: polarzing beam splitter cube , WP: waveplates, M: mirror. All but the \textit{BB1} sequence were tested with this arrangement. To test the \textit{BB1} sequence, we silightly changed the setup and placed the PBS
and the powermeter behind the waveplates.\label{fig:setup}}
\end{figure}

We guided the radiation from the various lasers to the test setup by optical fibers. We collimated the radiation by an adjustable collimation lens to a diameter of about 3~mm (FWHM). We polarized the light in the horizontal ($\mathcal{H}$) direction with a polarizer (P).
Afterwards the light propagated over a distance of 3~meters, and was reflected back by a mirror (M) close to normal incidence.
The angle separation between incoming and outgoing beam was less than 0.2$^\circ$.
The composite waveplate sequences under investigation were placed right in front of the mirror.
After propagating back by another 3 meters through the waveplate sequences, we guided the light onto a polarizing beam splitter (PBS) cube for polarization analysis.
We applied appropriate AR-coated PBSs matched to each wavelength applied in the tests.
We measured the transmitted and/or reflected power after the PBS with a suitable powermeter (PM).

The symmetric arrangement of rotation angles in the composite sequences with respect to the center waveplate (compare Table~\ref{tbl:data}, with the exception of the \textit{BB1} sequence for comparison) permitted us to implement a simple double-pass setup with back-reflection.
Thus, we replaced a setup intended for $N=7$ individual HWPs by 3 HWPs [H], 1 QWP [Q] and a retro-reflecting mirror [M], in order [HHHQM] (from left to right, compare Fig.~\ref{fig:setup}). This yields a much more compact setup and less optical elements are required. Only the \textit{BB1} sequence (with an asymmetric arrangement of angles) consisting of 5 HWPs was tested with a single-pass geometry.

We then aligned the waveplates such as to produce a rotation of the polarization by 90$^\circ$, i.e., from $\mathcal{H}$ to $\mathcal{V}$ polarization. For the tests of the QWP sequences, we passed the light twice through the \textit{complete} sequences (in order [QHHQM]), hence using them as HWPs. Also in this case, the waveplates were adjusted such as to produce a rotation of the polarization from $\mathcal{H}$ to $\mathcal{V}$. This simplified the measurements as only the transmitted or reflected power after the PBS had to be measured.

To align the rotation of the individual plates, we proceeded as follows: We started by setting the angles according to the calculated values within an accuracy of $\pm1^\circ$.
Next, we tilted the plates such that reflexes produced by the waveplates (e.g., at 532~nm, outside the bandwidth of the AR coating) did not coincide with the original beam to perturb our power measurement. As the conversion efficiency from $\mathcal{H}$ to $\mathcal{V}$ polarized light should reach a maximum at the center wavelength of 780~nm, we then measured the transmitted power at this wavelength behind the PBS ($\mathcal{H}$ polarization). Afterwards, we adjusted the optical axis of each plate iteratively within approximately $\pm1^\circ$ to minimize the transmission. With such alignment of the sequences, we then measured the conversion efficiencies from $\mathcal{H}$ to $\mathcal{V}$ polarization for all available wavelengths.
The conversion efficiency $I$ was measured asthe ratio of reflected power $P_{r}$ after the PBS compared to the incident power $P_0$ before the PBS.
In addition, we also considered transmission and reflection losses at the PBS.

\section{Results \& Discussion}\label{sec:results}

\subsection{Ultra-broadband Retarders}
\begin{figure}[ht]
\centerline{
\includegraphics[width=7.3cm]{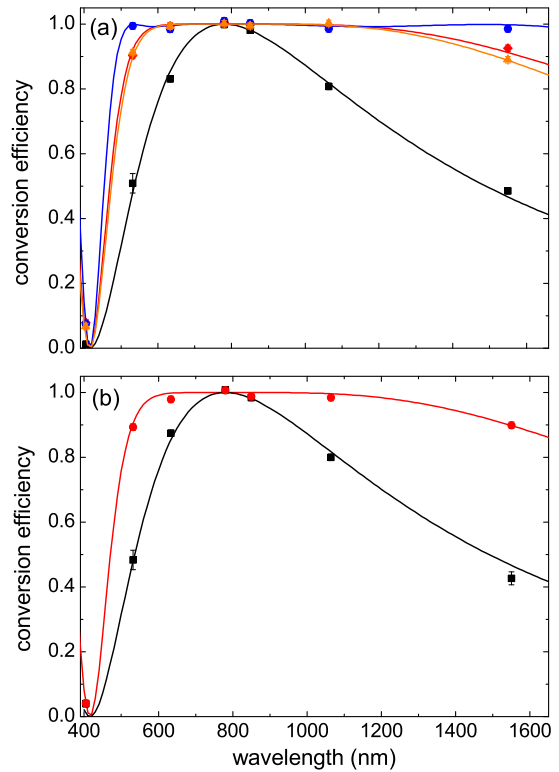}}
 \caption{(Color online) Measured (symbols) and calculated (lines) conversion efficiencies for several composite broadband HWP (a) and composite broadband QWP sequences (b). (a) The sequences are: single HWP [H] (black, squares); broadband HWPs (orange, triangles; red, rhombs; blue, dots; see Table~\ref{tbl:exp} for details). (b) single QWP in double-pass configuration [QM] (black, squares); broadband QWP in double-pass configuration (red, dots; see Table~\ref{tbl:exp} for details).
\label{fig:experiment}}
\end{figure}

Figure~\ref{fig:experiment}(a) shows the experimental data (symbols) of the conversion efficiency from $\mathcal{H}$ to $\mathcal{V}$ polarization for several composite broadband half-wave retarders in a broad wavelength range. The rotation angles of the optical axes of the single plates and their configuration for the various sequences are given in Table~\ref{tbl:exp}. These angles differ from the ones in Table~\ref{tbl:data}. We chose sequences with lower degree of retardation purity for the experimental demonstration in order to observe clear differences between them within our experimental precision. We therefore chose to present sequences with smaller, respectively, larger relative retardation variations $\Delta$, in order to demonstrate the functionality of our numerical optimization method which allows for arbitrary retardation specifications. For each set of data in Fig.~\ref{fig:experiment} we also show the theoretical prediction (solid lines) for the composite retarders based on the Jones calculus. Here, we also took the wavelength dependence $n_0(\lambda)-n_e(\lambda)$ \cite{G99} of the birefringence in the single quartz wave plates into account. The calculated conversion efficiency $I$ of the composite retarders from horizontal ($\mathcal{H}$) to vertical ($\mathcal{V}$) linear polarization is defined as follows : For a composite HWP we define $I=\left|\jones_{12}\right|^2$, while for a composite QWP we define $I=\left|(\jones^2)_{12}\right|^2$. The idea of this definition is, that in our experiment we test QWPs as HWPs by passing the light twice through the QWP sequences. Labels 1 and 2 denote the polarization basis vectors. The same symmetry as for $\delta(\varphi)$ relative to $\varphi=\ret0$ (see Sec.~\ref{sec:theory}) holds for the conversion efficiency profiles $I(\varphi)$ (for quarter-wave retarders $\theta_1=\pi/4$ is required).

For comparison of our new composite sequences, we also show the measured conversion efficiency of a conventional single HWP (black data, squares) as well as the \textit{BB1} sequence proposed by Ardavan \cite{A07b} (orange data, triangles). The \textit{BB1} sequence which is always implemented with $N=5$ individual HWPs already offers a clear improvement compared to the single HWP (black data, squares).

We consider now some of our proposed CP sequences with $N=7$ effective plates, implemented with $n=4$ individual plates. The data set in red rhombs shows the results for a new broadband sequence with a flat plateau.
The sequence has a slightly broader bandwidth of 700~nm compared to the \textit{BB1} sequence (orange data, triangles). This is partially due to the larger number ofeffective plates. 
The blue colored data (dots)show the results for a sequence where we allowed a deviation from the target retardation of about 10$^\circ$. This results in a $\mathcal{H}$ to $\mathcal{V}$ conversion efficiency of better than 0.99 over a huge bandwidth of more than 1100~nm. We note, that the performance of our composite broadband retarders is intrinsically limited by the bandwidth of the AR coatings. To our knowledge, AR bandwidths as large as the this retardation bandwidth are commercially not available. For the calibration of our data we therefore took the reflection losses at the stack of wave plates into account.

\begin{table}[ht]
	\caption{Angles of the individual wave plates for the experimentally tested sequences as shown in Fig.~\ref{fig:experiment} and
    Fig.~\ref{fig:experimentNB} in degree. For the broadband retarders, $\Delta I$ is the largest deviation of the conversion efficiency $I$ from unity in the design retardation range, while for narrowband retarders it is the largest deviation of $I$ from zero outside
    the design retardation range.}
	\label{tbl:exp}
	\begin{tabular}{ccc} \\ \hline\hline

	\multicolumn{3}{c}{(i) broadband half-wave retarders (see Fig.~\ref{fig:experiment}(a))}\\ \hline\hline
	line color & $\Delta I$ & ($\theta_1,\ldots,\theta_{n}$), [plate sequence] \\\hline
  orange & 0 & (0, 127.8, 23.3, 23.3, 127.8), [HHHHH] (\textit{BB1})\\
	red & 0 & (0, 106.0, 177.2, 232.4), [HHHQM]\\
	blue & $0.008$ & (-39.7, -23.3, 10.9, 78.7), [HHHQM]\\\hline\\\hline\hline

	\multicolumn{3}{c}{(ii) broadband quarter-wave retarders (see Fig.~\ref{fig:experiment}(b))}\\ \hline\hline
	line color & $\Delta I $ & ($\theta_1,\ldots,\theta_{n}$), [plate sequence] \\\hline
	red & 0 & (45.0, 96.8, 171.1, 96.8, 45.0), [QHHHQM]\\\hline \\\hline\hline

	\multicolumn{3}{c}{(iii) narrowband half-wave retarders (see Fig.~\ref{fig:experimentNB})}\\ \hline\hline
	line color & $\Delta I $ & ($\theta_1,\ldots,\theta_{n}$), [plate sequence]\\\hline
	magenta & 0 & (26.3, 7.2, 45.5, 14.6), [HHHQM]\\
  dark cyan & 0.0025 & (22.5, 34.7, 54.4, 55.8), [HHHQM]\\\hline
	\end{tabular}
\end{table}

Let us turn to an example for a novel composite QWP sequence. Figure~\ref{fig:experiment}(b) shows experimental data (symbols) along with theoretical predictions (lines). We tested the performance of the QWPs by passing the light beam twice through the arrangement, i.e., using the QWP in double-pass as HWPs.
As before, we investigated the performance of polarization rotation from $\mathcal{H}$ to $\mathcal{V}$ to define the conversion efficiency. For comparison, we
also recorded data for a single conventional zero-order QWP (black squares in Fig.~\ref{fig:experiment}(b)). The data for the new broadband composite QWP
are shown as red dots in the figure. As for the HWPs, we find good agreement between theoretical prediction and experiment.
The composite QWP offers a bandwidth of more than 500~nm where the conversion efficiency is above 0.99, i.e., well beyond that of a conventional zero-order plate.

We note the following issues: (i) We attribute the small deviations of some of the experimental data from the theoretical predictions to the effect of interferences due to multiple reflections at wavelengths outside the window of the AR coatings (see also Clarke \cite{C04c}).
(ii) In a simple picture, we would expect to find a minimum of the conversion efficiency at half of the design wavelength, hence at 390~nm. According to the calculation, this minimum occurs at an wavelength of about 415~nm. The shift is due to the wavelength dependence of the
refractive indices $n_{o,e}(\lambda)$.

\subsection{Narrowband Retarders}
Finally, we consider now the data taken for composite narrowband retarders. In the experiment, we applied the new narrowband retarders as highly chromatic HWPs between two crossed polarizers. Due to the symmetry of the rotation angles, we applied again a double-pass configuration. Hence we used $n=4$ plates in order [HHHQM] to implement a composite sequence of effectively $N=7$ plates. This is similar to a folded (type I) $\check{\mathrm{S}}$olc filter \cite{S65}.
\begin{figure}[ht]
\centerline{
\includegraphics[width=7.3cm]{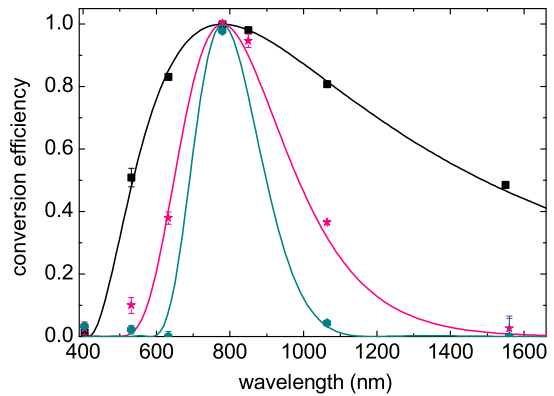}}
 \caption{(Color online) Measured (symbols)and calculated (lines) conversion efficiencies for several composite narrowband HWP sequences. The sequences are: single HWP [H] (black, squares); narrowband HWPs (magenta, stars; dark cyan, dots; see Table~\ref{tbl:exp} for details).
\label{fig:experimentNB}}
\end{figure}

The data sets in magenta stars and dark cyan dots in Fig.~\ref{fig:experimentNB} show the performances of two narrowband sequences.
The broader sequence (magenta) shows almost negligible sidebands at a conversion efficiency deviation below $\Delta I=10^{-4}$. 
The narrower sequence exhibits sidebands at a level below $\Delta I =2.5\times10^{-3}$. The composite narrowband retarders reduce the bandwidth by a factor of 2-3 compared to
a single conventional zero-order HWP (black data, squares). Thus, the filter performance significantly increases. We can get even smaller bandwidths for the narrowband sequences
by allowing larger deviations $\Delta$from the desired minimal retardation or by using more individual plates. We note, that the filter depicted by the data set in cyan dots exhibits a slightly larger bandwidth than the corresponding $\check{\mathrm{S}}$olc filter with 7 plates. However, the sidebands of this composite narrowband retarder are approx. 7 times smaller.

\section{Summary}

We presented new sequences of spectrally ultra-broadband and narrow-band polarization retarders. The concept is based on combination of conventional, identical zero-order waveplates with the optical axis of the single retarders rotated by appropriate angles. We used the mathematical analogy between the description of
polarization rotation and the dynamics of a light-driven two-level quantum system (i.e., the concept of composite pulses) to determine optimal rotation sequences of
waveplates. In comparison to single waveplates, the novel composite sequences can either reduce the chromatic dependence of the retardation
(i.e., to provide operation at ultra-broad bandwidth) or increase the chromatic dependence (i.e., to provide narrow-band retarders in filters). We theoretically
determined appropriate sequences of rotation angles for ultra-broadband and narrowband HWPs as well as QWPs. We tested some sequences experimentally
and compared their performance to conventional zero-order waveplates as well as (in the case of broadband HWPs) to a sequence known from the field of NMR. Our new
composite sequences proved to be superior in all cases. In particular, we demonstrated a composite HWP with a conversion efficiency from $\mathcal{H}$ to $\mathcal{V}$ polarized light above 0.99 in a huge spectral range with bandwidth of more than 1100~nm (centered at 780~nm). We also obtained a large increase of operation bandwidth for a novel composite QWP. Moreover, we found and demonstrated new sequences for composite highly-chromatic narrowband HWPs, which enables high-performance applications in filters such as Lyot or $\check{\mathrm{S}}$olc filters.

\section{Acknowledgments}
The research leading to these results has received funding from the Deutsche Forschungsgemeinschaft and the European Union Seventh Framework Programme (FP7/2007-2013) under grant agreements n$^\circ$ PCIG09-GA-2011-289305 and iQIT and the Bulgarian NSF grant DMU-03/103.

\end{document}